\documentclass[aps, prl, showpacs, showkeys, amsmath, amssymb, twocolumn,floatfix,superscriptaddress,nofootinbib]{revtex4-2}

\usepackage[colorlinks=true, linkcolor=blue, citecolor=blue, urlcolor=blue, linktoc=page, bookmarks=false, pdfstartview={FitH}, pdfborder={0 0 0.0 [3 3]}]{hyperref}
\usepackage[utf8]{inputenc}
\usepackage{xcolor}
\usepackage{soul}
\usepackage{amsmath,amssymb}
\usepackage{tensor} 
\usepackage{textgreek}
\usepackage{stmaryrd} 
\usepackage{graphicx}
\usepackage{subfiles}
\newcommand{\onlyinsubfile}[1]{#1}
\newcommand{\notinsubfile}[1]{}

\usepackage{tikz}
\usepackage[uncertainty-mode=separate]{siunitx}

\usepackage{todonotes}
\usepackage{xspace}
\usepackage{multirow}
\usepackage{booktabs}
\usepackage[normalem]{ulem} 

\newcommand*{\anue}{\ensuremath{\overline{\nu}_e}\xspace}
\newcommand*{\SinSqDouble}[2]{\ensuremath{\sin^22\theta_{#1#2}}}
\newcommand*{\SinSq}[2]{\ensuremath{\sin^2\theta_{#1#2}}}
\newcommand*{\CosSq}[2]{\ensuremath{\cos^2\theta_{#1#2}}}
\newcommand*{\CosTetr}[2]{\ensuremath{\cos^4\theta_{#1#2}}}
\newcommand*{\Dm}[2]{\ensuremath{\Delta m^2_{#1#2}}}
\newcommand*{\mtheta}[2]{\ensuremath{\theta_{#1#2}}}

\newcommand*{\CLs}{CL$_{\rm s}$\xspace}
\newcommand*{\eV}[1]{\text{eV}\ensuremath{^{#1}}}

\begin{document}
\renewcommand{\onlyinsubfile}[1]{}
\renewcommand{\notinsubfile}[1]{#1}

\title{Search for a sub-eV sterile neutrino using Daya Bay's full dataset}
\newcommand{\NCSL}{\affiliation{New Cornerstone Science Laboratory, Institute of High Energy Physics, Beijing}}
\newcommand{\IHEP}{\affiliation{Institute~of~High~Energy~Physics, Beijing}}
\newcommand{\Wisconsin}{\affiliation{University~of~Wisconsin, Madison, Wisconsin 53706}}
\newcommand{\Yale}{\affiliation{Wright~Laboratory and Department~of~Physics, Yale~University, New~Haven, Connecticut 06520}} 
\newcommand{\BNL}{\affiliation{Brookhaven~National~Laboratory, Upton, New York 11973}}
\newcommand{\NTU}{\affiliation{Department of Physics, National~Taiwan~University, Taipei}}
\newcommand{\NUU}{\affiliation{National~United~University, Miao-Li}}
\newcommand{\Dubna}{\affiliation{Joint~Institute~for~Nuclear~Research, Dubna, Moscow~Region}}
\newcommand{\CalTech}{\affiliation{California~Institute~of~Technology, Pasadena, California 91125}}
\newcommand{\CUHK}{\affiliation{Chinese~University~of~Hong~Kong, Hong~Kong}}
\newcommand{\NCTU}{\affiliation{Institute~of~Physics, National~Chiao-Tung~University, Hsinchu}}
\newcommand{\NJU}{\affiliation{Nanjing~University, Nanjing}}
\newcommand{\TsingHua}{\affiliation{Department~of~Engineering~Physics, Tsinghua~University, Beijing}}
\newcommand{\SZU}{\affiliation{Shenzhen~University, Shenzhen}}
\newcommand{\NCEPU}{\affiliation{North~China~Electric~Power~University, Beijing}}
\newcommand{\Siena}{\affiliation{Siena~College, Loudonville, New York  12211}}
\newcommand{\IIT}{\affiliation{Department of Physics, Illinois~Institute~of~Technology, Chicago, Illinois  60616}}
\newcommand{\LBNL}{\affiliation{Lawrence~Berkeley~National~Laboratory, Berkeley, California 94720}}
\newcommand{\UIUC}{\affiliation{Department of Physics, University~of~Illinois~at~Urbana-Champaign, Urbana, Illinois 61801}}
\newcommand{\SJTU}{\affiliation{Department of Physics and Astronomy, Shanghai Jiao Tong University, Shanghai Laboratory for Particle Physics and Cosmology, Shanghai}}
\newcommand{\BNU}{\affiliation{Beijing~Normal~University, Beijing}}
\newcommand{\WM}{\affiliation{College~of~William~and~Mary, Williamsburg, Virginia  23187}}
\newcommand{\Princeton}{\affiliation{Joseph Henry Laboratories, Princeton~University, Princeton, New~Jersey 08544}}
\newcommand{\VirginiaTech}{\affiliation{Center for Neutrino Physics, Virginia~Tech, Blacksburg, Virginia  24061}}
\newcommand{\CIAE}{\affiliation{China~Institute~of~Atomic~Energy, Beijing}}
\newcommand{\SDU}{\affiliation{Shandong~University, Jinan}}
\newcommand{\NanKai}{\affiliation{School of Physics, Nankai~University, Tianjin}}
\newcommand{\UC}{\affiliation{Department of Physics, University~of~Cincinnati, Cincinnati, Ohio 45221}}
\newcommand{\DGUT}{\affiliation{Dongguan~University~of~Technology, Dongguan}}
\newcommand{\XJTU}{\affiliation{Department of Nuclear Science and Technology, School of Energy and Power Engineering, Xi'an Jiaotong University, Xi'an}}
\newcommand{\UCB}{\affiliation{Department of Physics, University~of~California, Berkeley, California  94720}}
\newcommand{\HKU}{\affiliation{Department of Physics, The~University~of~Hong~Kong, Pokfulam, Hong~Kong}}
\newcommand{\Charles}{\affiliation{Charles~University, Faculty~of~Mathematics~and~Physics, Prague}} 
\newcommand{\USTC}{\affiliation{University~of~Science~and~Technology~of~China, Hefei}}
\newcommand{\TempleUniversity}{\affiliation{Department~of~Physics, College~of~Science~and~Technology, Temple~University, Philadelphia, Pennsylvania  19122}}
\newcommand{\CGNPG}{\affiliation{China General Nuclear Power Group, Shenzhen}}
\newcommand{\NUDT}{\affiliation{College of Electronic Science and Engineering, National University of Defense Technology, Changsha}} 
\newcommand{\IowaState}{\affiliation{Iowa~State~University, Ames, Iowa  50011}}
\newcommand{\ZSU}{\affiliation{Sun Yat-Sen (Zhongshan) University, Guangzhou}}
\newcommand{\CQU}{\affiliation{Chongqing University, Chongqing}} 
\newcommand{\BCC}{\altaffiliation[Now at ]{Department of Chemistry and Chemical Technology, Bronx Community College, Bronx, New York  10453}} 

\newcommand{\UCI}{\affiliation{Department of Physics and Astronomy, University of California, Irvine, California 92697}} 
\newcommand{\GXU}{\affiliation{Guangxi University, No.100 Daxue East Road, Nanning}} 
\newcommand{\HKUST}{\affiliation{The Hong Kong University of Science and Technology, Clear Water Bay, Hong Kong}} 
\newcommand{\Rochester}{\altaffiliation[Now at ]{Department of Physics and Astronomy, University of Rochester, Rochester, New York 14627}} 

\newcommand{\LSU}{\altaffiliation[Now at ]{Department of Physics and Astronomy, Louisiana State University, Baton Rouge, LA 70803}} 

\author{F.~P.~An}\ZSU
\author{W.~D.~Bai}\ZSU
\author{A.~B.~Balantekin}\Wisconsin
\author{M.~Bishai}\BNL
\author{S.~Blyth}\NTU
\author{G.~F.~Cao}\IHEP
\author{J.~Cao}\IHEP\NCSL
\author{J.~F.~Chang}\IHEP
\author{Y.~Chang}\NUU
\author{H.~S.~Chen}\IHEP
% ...

\collaboration{The Daya Bay Collaboration}\noaffiliation

\date{\today}

\begin{abstract}
  This Letter presents results of a search for the mixing of a sub-eV sterile neutrino with three active neutrinos based on the full data sample of the Daya Bay Reactor Neutrino Experiment, collected during 3158 days of detector operation, which contains $5.55 \times 10^{6}$ reactor \anue candidates identified as inverse beta-decay interactions followed by neutron-capture on gadolinium. The analysis benefits from a doubling of the statistics of our previous result and from improvements of several important systematic uncertainties. 
  No significant oscillation due to mixing of a sub-eV sterile neutrino with active neutrinos was found. Exclusion limits are set by both Feldman-Cousins and CLs methods.
  Light sterile neutrino mixing with $\sin^2 2\theta_{14} \gtrsim 0.01$ can be excluded at 95\% confidence level in the region of $0.01$ eV$^2 \lesssim |\Delta m^{2}_{41}| \lesssim 0.1 $ eV$^2$. This result represents the world-leading constraints in the region of  $2 \times 10^{-4}$ eV$^2 \lesssim |\Delta m^{2}_{41}| \lesssim 0.2 $ eV$^2$. 
\end{abstract}

\maketitle

The existence of neutrino oscillation has been firmly demonstrated by many experiments, ruling out the original Standard Model's postulate that all neutrino states are massless.
Although most of the findings agree with the hypothesis that there are three neutrino mass states, 
there exist a few anomalies~\cite{LSND:2001aii,Barinov:2022wfh} and indications~\cite{Serebrov:2020kmd} that may be explained
by the existence of extra neutrino mass states~\cite{Acero:2022wqg}.
Precision measurements of the $Z$-boson width are consistent with three light neutrino species
that participate in the weak interaction \cite{ALEPH:2005ab} so any additional neutrino species must be ``sterile", that is, not subject to the weak interaction.

As for the mass of an extra neutrino state, in theory any value is possible. A mass as large as $10^{15}$ GeV is considered by the seesaw mechanism, 
which can both generate the very light active-neutrino masses and produce the baryon asymmetry of the universe~\cite{seesaw1,seesaw2,seesaw3,seesaw4,seesaw5}. 
In contrast, a keV-range sterile neutrino is a possible candidate for warm dark matter \cite{Lovell:2020vlf}.
The most stringent limits on the mass of a light relativistic sterile neutrino again come from cosmology, which is sensitive to the sum of neutrino masses~\cite{Acero:2022wqg}.
The current limit of $\sum m_\nu<0.12$~eV at 95\% C.L.\cite{Planck:2018vyg} remains consistent with the existence of a sub-eV sterile neutrino; 
this mass constraint is loosened to a few MeV when sterile neutrino self-interactions are allowed~\cite{Corona:2021qxl}.

In the minimal ``3+1'' extension of the three-neutrino model, considering one sterile neutrino in addition to the three active neutrinos, 
the flavor eigenstates $\nu_\alpha$ ($\alpha=e,\mu,\tau,s$) are related to the four mass eigenstates $\nu_i$ as:
\begin{align}
    \nu_\alpha = \sum\limits_{i=1}^4 U_{\alpha i} \nu_i,
\end{align}
where $U$ is a unitary $4\times4$ mixing matrix, analogous to the Pontecorvo-Maki-Nakagawa-Sakata (PMNS) matrix from the three-neutrino scenario. 
The matrix is in general parameterized~\cite{ParticleDataGroup:2020ssz,An_2014} by six mixing angles $\theta_{ij}$ and three CP-violating phases $\delta_i$. 
The survival probability of electron antineutrinos ($P_{\anue\shortrightarrow\anue}$) is a function of the neutrino energy $E$ and the distance traveled (\emph{i.e.} baseline) $L$ as:
\begin{align}
    P_{\anue\shortrightarrow\anue} = 1 - 
    4 \sum\limits_{j>i}^4 
    \left|U_{e i}\right|^2
    \left|U_{e j}\right|^2
    \sin^2 \Delta_{ji},
    \label{eq:psur}
\end{align}
where $\Delta_{ji}\approx1.267 \Delta m^2_{ji}[\text{eV}^2]L\text{[m]}/E\text{[MeV]}$, and $\Delta m^{2}_{ji}=m_j^2-m_i^2$ is the mass-squared difference of the mass eigenstates $\nu_j$ and $\nu_i$. 
The survival probability of \anue depends only on three mixing angles \mtheta12, \mtheta13 and \mtheta14 and six mass-squared differences, 
only three of which are independent.
According to Eq. \eqref{eq:psur} the survival probability oscillates with wavelength proportional to $E/\Delta m^{2}$.
Assuming $m_1$ to be the lightest and $\Dm21 \ll \Dm41$, the survival probability may be approximated for baselines below hundreds of meters and for MeV-scale energies as\footnote{In this analysis, 
we use the full formula for neutrino oscillation shown in Eq.~\ref{eq:psur}.}:
\begin{equation}
\begin{aligned}
    P_{\anue \shortrightarrow \anue} &\approx \\
    1 &- 
    \SinSqDouble14(\CosSq13\sin^2\Delta_{41} + \SinSq13\sin^2\Delta_{43}) 
    \\
    &- \CosTetr14 \SinSqDouble13 \sin^2\Delta_{32}
   .\label{eq:psur_approx}
\end{aligned}
\end{equation}
One possible hint of a sterile neutrino would thus be a deficit of the electron-antineutrino event rate, accompanied by the dependence of the energy-spectrum distortion pattern on the distance between source and detector.

Such a deficit would be visible in the Daya Bay Reactor Neutrino Experiment with baselines spanning from 360~m to 1900~m~\cite{article:long_osc_2016} that can provide world-leading sensitivity to a sterile neutrino with $\left|\Dm41\right| < 0.2$ eV$^2$.
This Letter reports the search for such a state with the Daya Bay experiment's full data sample of inverse beta decay (IBD) interactions identified by subsequent neutron-capture on gadolinium.

From 2011 to 2020, the Daya Bay experiment operated with up to eight identically designed antineutrino detectors (ADs) near the Daya Bay and Ling Ao nuclear power plants 
in southern China ~\cite{DayaBay:2015kir,dayabaycollaboration2022precision}. The ADs were distributed among two Near experimental halls (EH1 and EH2, each with up to 2 ADs) and one Far hall (EH3, with up to 4 ADs). 
Each AD contained a target mass of $\sim$20~tons of gadolinium-doped liquid scintillator, observed by 192 8-inch photomultiplier tubes (PMTs).

As usual for reactor-based experiments, the Daya Bay experiment detects electron antineutrinos via the IBD interaction:
$\bar\nu_e + p \to e^+ + n$.
The outgoing positron ionizes the scintillator and annihilates with an electron, producing a signal corresponding to a ``prompt'' energy with
$E_p \approx E_\nu - \num{0.8}$~MeV.
The neutron is captured with an average delay of 28 $\mu$s, predominantly by gadolinium, which deexcites via a cascade of gamma rays totaling $\sim$\num{8}~MeV, forming the ``delayed" signal. 
In this channel, the close temporal coincidence of the two signals, together with the high energy of the delayed signal,
allow for an average background-to-signal ratio of $\sim1.5\%$.

The Daya Bay experiment operated with different detector  configurations in three consecutive periods, which we label the 6-AD, 8-AD, and 7-AD periods, based on the number of active ADs. 
In this analysis, we adopt the same IBD data sample used in the most recent three-flavor neutrino oscillation analysis~\cite{dayabaycollaboration2022precision}. 
This full dataset, with a total of $\sim$5.55 million IBD candidates, has twice the statistics of the sample used in the previous sterile neutrino search~\cite{article:sterile_2020}. 
The daily IBD rates and the estimated backgrounds in the three halls are summarized in Table~\ref{table_full_dataset} for the three periods. In addition to the increased statistics, 
this analysis benefits from improvements to four systematic uncertainties: 
i) The $^9$Li/$^8$He background was estimated using a new multi-dimensional fitting method~\cite{dayabaycollaboration2022precision}; 
ii) The effect of spent nuclear fuel was derived from detailed history on reactor operation~\cite{Ma:2015lsv,Adey_2018}; 
iii) The channel-wise electronics nonlinearity was recalibrated using an FADC readout system~\cite{P17B_nonlinearity,Adey_2018}; 
iv) The energy response model was constrained with new calibration data~\cite{P17B_nonlinearity,Adey_2018}.

The background rate is dominated by accidental coincidences of uncorrelated signals satisfying the IBD selection criteria. 
To mitigate this background, in the summer of 2012, the $^{241}\text{Am}\mbox{-}^{13}\text{C}$ neutron sources were removed from two of the automated calibration units on each far AD, 
halving the rate of delayed-like uncorrelated signals in the far hall~\cite{article:long_osc_2016}.
Although these accidentals dominate the background rate, the uncertainty of the total background rate is dominated by the contribution from correlated pairs induced primarily by cosmic-ray muons.
In particular, cosmogenic $^{9}$Li/$^{8}$He is a well-known background in liquid scintillation detectors used for reactor neutrino experiments, and in the Daya Bay experiment it is the leading contributor to the background uncertainty.
In this analysis, the relative uncertainty of the estimated $^{9}$Li/$^{8}$He background rate has been reduced from $\sim$35\%~\cite{article:long_osc_2016} to $<$25\% by 
taking into account the correlated temporal and energy information of the IBD candidates~\cite{dayabaycollaboration2022precision}.

\begin{table*}[t!]
\centering
\renewcommand{\arraystretch}{1.2}
  \caption{Summary of IBD signal and background for the full dataset for 3 detector configurations and 3 experimental halls (EHs). The errors include statistical and systematic uncertainties. 
  The effective live time is the product of live time, the efficiency of the muon veto and the efficiency of the multiplicity selection.}
\resizebox{\linewidth}{!}{
\begin{tabular}{l
    S[table-format=3.2+-1.2]
    S[table-format=3.2+-1.2]
    S[table-format=3.2+-1.2]
    c
    S[table-format=3.2+-1.2]
    S[table-format=3.2+-1.2]
    S[table-format=3.2+-1.2]
    c
    S[table-format=3.2+-1.2]
    S[table-format=3.2+-1.2]
    S[table-format=3.2+-1.2]
    @{\hspace*{3mm}}
    }

      \toprule\toprule
      \multirow{2.4}{*}{\phantom{xx}} & \multicolumn{3}{c}{6-AD} & \phantom{x} & \multicolumn{3}{c}{8-AD} & \phantom{x} & \multicolumn{3}{c}{7-AD} \\
      \multirow{2.4}{*}{\phantom{xx}} & \multicolumn{3}{c}{24 Dec 2011 - 28 Jul 2012} & \phantom{x} & \multicolumn{3}{c}{19 Oct 2012 - 20 Dec 2016} & \phantom{x} & \multicolumn{3}{c}{26 Jan 2017 - 12 Dec 2020}\\
      \cmidrule{2-4}
      \cmidrule{6-8}
      \cmidrule{10-12}
                            & \multicolumn{1}{c}{EH1}           & \multicolumn{1}{c}{EH2}         & \multicolumn{1}{c}{EH3} & & \multicolumn{1}{c}{EH1} & \multicolumn{1}{c}{EH2} & \multicolumn{1}{c}{EH3}  & & \multicolumn{1}{c}{EH1} & \multicolumn{1}{c}{EH2} & \multicolumn{1}{c}{EH3} \\
                            \midrule
      Number of active ADs                     & \num{2}           & \num{1}          & \num{3}          &  & \num{2}           & \num{2}          & \num{4}          &  & \num{1}           & \num{2}          & \num{4}          \\
      
      IBD candidates                    & \num{199 637}     & \num{92 011}     & \num{40 531}     &  & \num{1 400 061}   & \num{1325843}    & \num{389030}     &  & \num{637112}      & \num{1127040}    & \num{334853}     \\
      
      Effective live time [day]         & \num{148.987}     & \num{155.973}    & \num{178.056}    &  & \num{1037.910}    & \num{1095.292}   & \num{1284.509}   &  & \num{887.457}     & \num{934.005}    & \num{1095.468}   \\
      
      Background [/AD/day] & \num{12.36+-0.58} & \num{9.96+-0.59} & \num{3.16+-0.08} &  & \num{10.87+-0.57} & \num{8.26+-0.42} & \num{1.11+-0.03} &  & \num{10.61+-0.81} & \num{7.75+-0.41} & \num{0.94+-0.03} \\
      
      Signal [/AD/day]     & \num{657.6+-1.6}  & \num{580.0+-2.0} & \num{72.7+-0.4}  &  & \num{633.6+-0.8}  & \num{597.0+-0.7} & \num{74.6+-0.1}  &  & \num{707.3+-1.2}  & \num{595.6+-0.7} & \num{75.5+-0.1}  \\
      \bottomrule\bottomrule
    \end{tabular}
    }
  \label{table_full_dataset}
\end{table*}

The antineutrino flux is predicted from thermal-power data and from the calculated fission fractions of each fuel cycle. 
Uncertainties in the thermal-power data lead to a core-to-core uncorrelated flux uncertainty of $0.5\%$, 
while a 0.6\% uncorrelated uncertainty per core in the $\bar{\nu}_{e}$ yield was introduced by the uncertainties of the fission fractions.
The spent nuclear fuel in the cooling pools adjacent to each core contributes $0.3\pm0.1$\%
to the predicted neutrino flux~\cite{DayaBay:2016ssb,Ma:2015lsv}.
The correlated flux uncertainty
includes contributions from the 0.2\% uncertainties of the mean energy released per fission~\cite{DYB:fission_energy}.
However, the primary contributors to the correlated uncertainty are the theoretical uncertainties on forbidden decays 
and the missing information in published nuclear data tables~\cite{Hayes:2013wra}, which lead to a conservative total of 5\% for the core-to-core correlated flux uncertainty.
The size of the reactor cores and ADs has negligible impact on the  sterile neutrino sensitivity due to the relatively long baselines.

The nominal reactor antineutrino spectra from \nuclide[235]{U}, \nuclide[239]{Pu}, \nuclide[241]{Pu} and \nuclide[238]{U} 
were predicted using the models of Huber~\cite{model:Huber} and Mueller \emph{et al.}~\cite{model:Mueller}. 
Multiple methods~\cite{article:long_osc_2016} were used to account for the known disagreements between these models and existing measurements~\cite{DayaBay:2019yxq}. 
One method applied enlarged uncertainties, ranging from 10\% to 40\% as a function of energy, to the neutrino spectra.
An alternative method used the observations of the near detectors to predict the observations of the far detectors, while another used free parameters to modify the predicted antineutrino spectra in the fit. As an additional cross-check, the analysis was also repeated using summation spectra~\cite{model:Summation2018}. 
All of the aforementioned methods produce consistent results.

To predict the IBD rate and reconstructed prompt-energy spectrum at each AD, the reactor spectra were multiplied by the neutrino oscillation probability (Eq.~\eqref{eq:psur}), the IBD differential cross section,
the number of target protons, the detection efficiency, and the detector's energy response, which maps positron energy to reconstructed prompt energy. The energy response model
considers energy resolution and nonlinearity and possible energy loss in the inner acrylic vessel.

To further improve the energy nonlinearity model, in December 2015 a full flash ADC (FADC) readout system was installed in EH1-AD1, recording PMT waveforms at 1~GHz and 10-bit resolution~\cite{P17B_nonlinearity}. 
A deconvolution method was applied to the waveforms to minimize any dependence on the single-photoelectron pulse shape (in particular the overshoot) and to extract the integrated charge with minimal bias. 
The residual nonlinearity in the corrected charge from a single waveform was estimated to be less than 1\%, resulting in a 0.2\%  nonlinearity in the total charge measurement for each event~\cite{P17B_nonlinearity}.
In addition, a special calibration campaign in January 2017 improved the knowledge of optical shadowing by the radioactive source enclosures, 
reducing the energy calibration uncertainties of $\gamma$ rays from 1\% to 0.5\% \cite{P17B_nonlinearity}. 
This combination of improvements, when applied to calibration data, led to a significant improvement in the energy response model for positrons. 
The uncertainty improved from $\sim1\%$~\cite{article:sterile_2020,article:long_osc_2016} to $< 0.5\%$~\cite{P17B_nonlinearity,Adey_2018} for $E_{p} > 2$ MeV.

\begin{figure}[htb]
  \centering
  \includegraphics[scale=0.72]{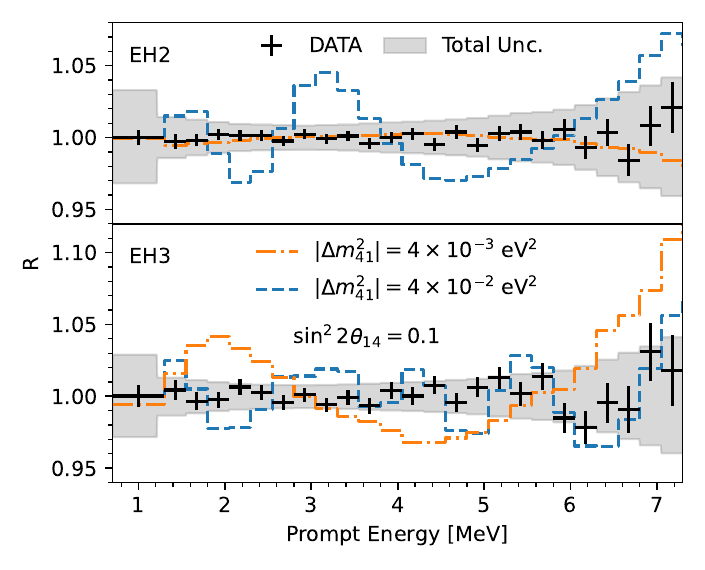}
  \caption{Impact of sterile neutrino oscillation on the double ratios $(R)$ of the spectra, obtained at EH2(3), to EH1 as 
  $R=(\text{M}_{\text{EH2(3)}}/\text{P}^{3\nu}_{\text {EH2(3)}})/(\text{M}_{\text{EH1}}/\text{P}^{3\nu}_{\text{EH1}})$, 
  where M is the measured prompt energy spectrum, and $\text{P}^{3\nu}$ is the three-neutrino prediction with the best-fit parameters. 
  The error bands represent the total uncertainties (statistical and systematic) of the three-neutrino prediction. 
  The error bars represent only the statistical uncertainties. For the predictions, $\sin^{2}2\theta_{14}=0.1$ and two $\left|\Dm41\right|$ values are shown.
  {\hspace*{\fill} }
  }
  \label{spectra_by_EH1_ratio}
\end{figure} 

In the region of $\left|\Dm41\right|$ where the Daya Bay experiment provides world-leading sensitivity on $\sin^2 2\theta_{14}$, 
the bulk of the sensitivity comes from relative measurements between the experimental halls.
Figure~\ref{spectra_by_EH1_ratio} shows the double ratio of the measured to three-neutrino-predicted spectra of EH2 and EH3 to that of EH1.
The data are compared to two four-neutrino predictions. 
The data are well contained in the uncertainty band, indicating that the data are consistent with the three-neutrino prediction.  

To quantify our measurement, several different statistical methods have been used in this analysis, yielding consistent results. 
One method used a $\chi^{2}$ statistic based on the binned maximum likelihood ratio with systematic uncertainties treated via Gaussian nuisance terms. 
An alternative method built a $\chi^2$ function using a covariance matrix generated by a toy Monte Carlo that includes fluctuations due to systematic uncertainties. 
Hybrids of the two approaches have also been tested. The free parameters are $\sin^2 2\theta_{13}$, $\sin^2 2\theta_{14}$ and $\Delta m^2_{41}$.
We used $\Delta m^2_{32}$ = $\num{2.453(0.034)e-3}$ \eV2 \cite{ParticleDataGroup:2020ssz} (normal mass ordering).

To test the consistency between the Daya Bay experimental data and the three- ($3\nu$) or four-neutrino ($4\nu$) hypotheses, 
we first calculated $\chi^2(\eta)$ at each point in the $\eta \equiv \left(\sin^2 2\theta_{14}, \left|\Dm41\right|\right)$ parameter space by profiling over $\sin^2 2\theta_{13}$ and the nuisance parameters. 
These parameters are those which minimize $\chi^2$ at each point in $\eta$. 
We then defined a test statistic $\Delta \chi^{2} = \chi^2(\eta_{(0, 0)}) - \chi^2({\eta}_{\text{bf}})$ with 2 degrees of freedom $\left(\sin^2 2\theta_{14}, \left|\Dm41\right|\right)$, 
where $\eta_{(0, 0)}$ represents the $3\nu$ oscillation assumption  and ${\eta}_{\text{bf}}$ represents the global best fit under the assumption of $4\nu$ oscillation. 
The Daya Bay experimental data gave $\Delta \chi^{2} = $ 2.3, corresponding to a $p$-value~\cite{ParticleDataGroup:2020ssz} of $0.86$, 
obtained from the $\Delta \chi^{2}$ distribution generated by Monte Carlo samples under the $3\nu$ oscillation hypothesis including statistical and systematic variations. 
This indicates that no significant signal of a sterile neutrino was observed.

The Feldman-Cousins (FC) method~\cite{article:FC_1998_Feldman} was used to set confidence intervals in the $\eta$  parameter space.
For each point ${\eta}$, a distribution of $\Delta\chi^2 =  \chi^2({\eta}) - \chi^2({\eta}_{\text{bf}})$ was generated from 1000 pseudoexperiments with both statistical and systematic variations considered. 
Based on the $\Delta\chi^2$ distribution and the $\Delta\chi^2$ observed with the data, a $p$-value for each $\eta$ point was calculated. 
The $1-\alpha$ confidence interval boundary was set where  $p$-value=$\alpha$.

An alternative method for determining limits is the Gaussian \CLs method~\cite{article:CLs_Gaussian_2016_Qian} 
based on a two-hypothesis test, comparing the null hypothesis ($3\nu$) and the alternative hypothesis ($4\nu$) for each point $\eta$. 
Using the resulting test statistic $\Delta \chi^{2} = \chi^2(\eta) - \chi^2(\eta_{(0, 0)})$,
the corresponding \CLs value was calculated~\cite{ParticleDataGroup:2020ssz}:
\begin{equation}
    \text{CL$_{\rm s}$} = \frac{p_{4\nu}}{1-p_{3\nu}},
\end{equation}
where $p_{4\nu}$ and $p_{3\nu}$ are the $p$-values of the two hypothesis. The condition of \CLs $< \alpha$ was used to set the $(1-\alpha)$ \CLs exclusion region~\cite{ParticleDataGroup:2020ssz}.

\begin{figure}[h]
    \centering
	\begin{tikzpicture}
	\node at (0., 0) {\includegraphics[scale=0.72]{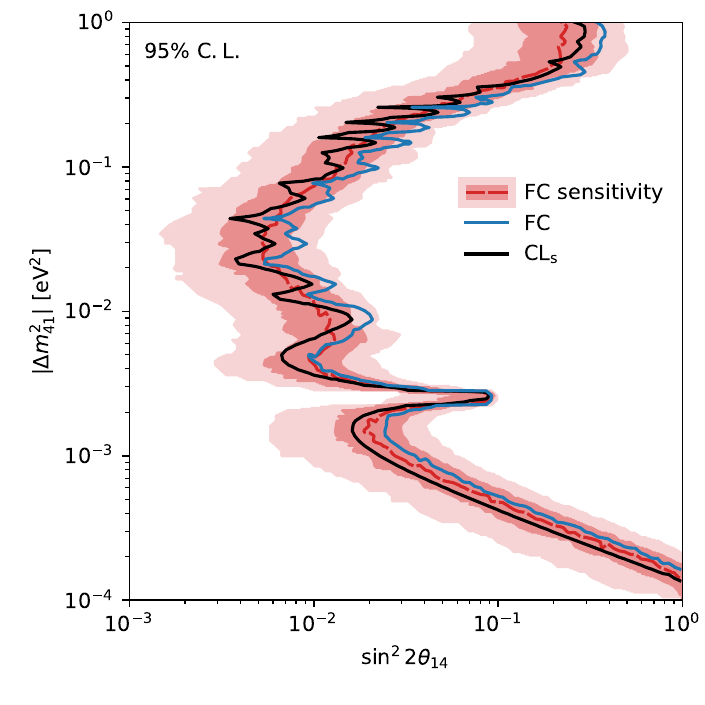}};
	\end{tikzpicture}
	\caption{
	 Exclusion contours at 95\% C.L., obtained by the FC and \CLs methods.
	 For the FC method the 1$\sigma$ and 2$\sigma$ bands are also shown, which account for the statistical and all systematic uncertainties.
	 The parameter space to the right of the contours is excluded.
	}
	\label{fig:CLs_FC_plot_excluded}
\end{figure}

The results of the application of the FC and the \CLs methods are shown in Figure~\ref{fig:CLs_FC_plot_excluded}.   
The acronym ``95\% C.L.'' represents both the 95\% confidence interval for the FC method and the 95\% exclusion region for the \CLs method.
Both contours show consistent features. The \CLs method provides more stringent limits due to the different definition of the test statistic $\Delta\chi^2$. 
The decrease of sensitivity in the region of $\left|\Dm41\right|$ $\approx\Delta m^2_{32}$ is related to the fact that the oscillations to the sterile neutrino state are not distinguishable from $3\nu$ oscillations.
At the baseline of the Daya Bay experiment, the choice of neutrino mass ordering has a marginal impact on the results.
The best sensitivity to $\sin^2 2\theta_{14}$ is achieved in the region $10^{-2}$~eV$^2 \lesssim \left|\Dm41\right| \lesssim 0.1$ eV$^2$ where the measurement relies on the relative spectral differences 
between the detectors of the Near and Far experimental halls. For the higher mass-squared difference, the sensitivity decreases as the detectors become insensitive to the shape of sterile-active oscillations. 
The contour tends to a constant in the region $|\Delta m^2_{41}| \gtrsim 0.5$~eV$^2$ where the prediction becomes limited by the uncertainty in the reactor antineutrino flux.

\begin{figure}
    \centering
	\begin{tikzpicture}
	\node at (0., 0.) {\includegraphics[scale=0.72]{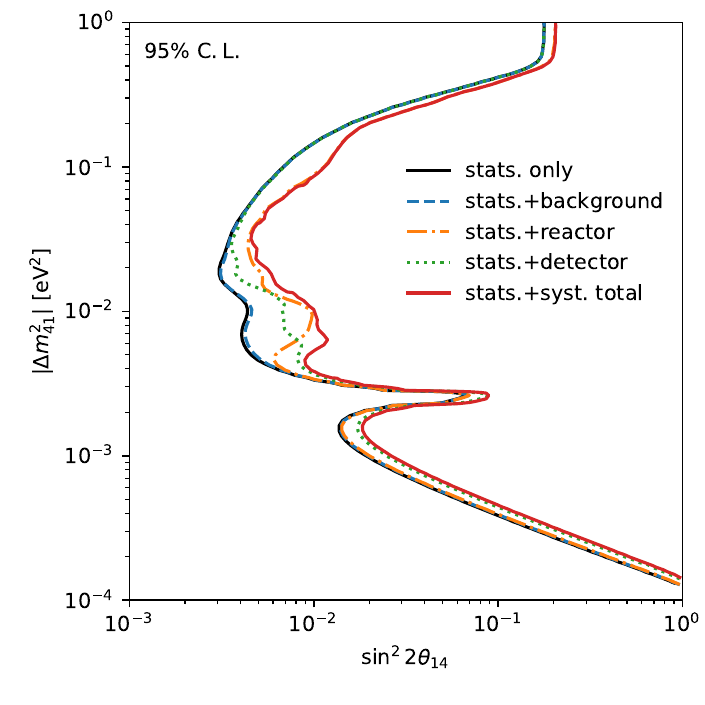}};
	\end{tikzpicture}
	\caption{
    Sensitivity contours at 95\% C.L. with the \CLs method for different combinations of statistical and systematic uncertainties. The systematic uncertainties are split into three groups: background, reactor, and detector.}
	\label{fig:CLs_plot_systematic}
\end{figure}

To demonstrate the effects of different types of uncertainties~\cite{article:long_osc_2016}, the sensitivity with the \CLs method under various scenarios is shown in Figure~\ref{fig:CLs_plot_systematic}.
The main effect of the reactor antineutrino flux uncertainties falls in the region of $|\Delta m^2_{41}| \gtrsim \num{4e-3}$~eV$^2$, 
where the sensitivity is dominated by the relative spectral difference between the two Near EHs~\cite{An_2014}.  
The sensitivity for $\left|\Dm41\right| \lesssim \num{3e-2}$~eV$^2$ is affected by the uncertainties of the detector energy response model, 
where the relative difference between EH3 and the two Near EHs plays the most important role~\cite{An_2014}. 
Background uncertainties have a negligible effect on the sensitivity contour due to both the low background level and the accurate estimation of the background. 

\begin{figure}[h]
    \centering
	\begin{tikzpicture}
    	\node at (0., 0) {\includegraphics[scale=0.72]{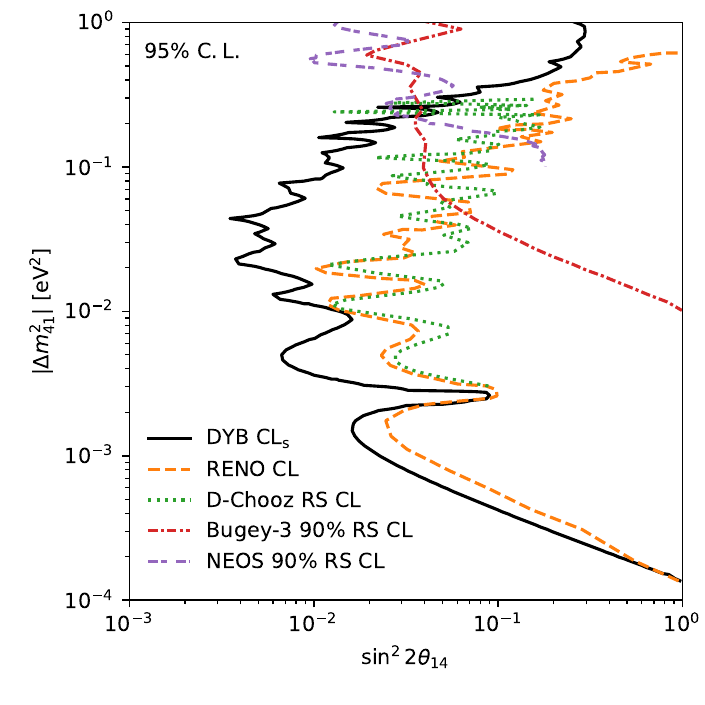}};
	\end{tikzpicture}
	\caption{
	Comparison of the Daya Bay experiment's results with those from other reactor experiments. The methods used to generate these contours are not identical. 
    The Daya Bay experiment (DYB) used the \CLs method, while RENO used traditional method based on Wilk's theorem~\cite{RENO_paper,Wilks:1938dza}, 
    and Double Chooz (D-Chooz)~\cite{DoubleChooz:2020pnv}, Bugey-3~\cite{Bugey_paper} and NEOS~\cite{NEOS_paper} all used the raster scan (RS)~\cite{Lyons:2014kta} method. 
    The contours from applying the RS and traditional method to the Daya Bay data are shown in the supplemental material, and are consistent with the contour from the \CLs method.
	}
	\label{fig:CLs_FC_others}
\end{figure}

A comparison with the results of other short and medium baseline reactor neutrino experiments is shown in Figure~\ref{fig:CLs_FC_others}. 
The Day Bay experiment is able to set the most stringent upper limits on light sterile-active neutrino mixing for $2 \times 10^{-4}$ eV$^2 \lesssim \left|\Dm41\right| \lesssim 0.2$ eV$^2$ 
due to its high statistics and well-controlled systematic uncertainties.
Currently, short baseline ($\lesssim$ 100 m) experiments like Bugey-3 and NEOS give more stringent limits than Daya Bay in the region of $\left|\Dm41\right| \gtrsim 0.2$ eV$^{2}$.
In the future, the JUNO experiment will dominate in the region of $\left|\Dm41\right| \lesssim 10^{-3}$ eV$^{2}$, because of its long baseline ($\sim 53$ km) and  superb energy resolution~\cite{JUNO:yellowbook}.

In summary, we report the results of a light sterile neutrino search using the full dataset of the Daya Bay Reactor Neutrino Experiment. 
The data are consistent with the canonical extension of the Standard Model with three massive neutrinos. 
No evidence of a light sterile neutrino was found. 
The world's most stringent limits on the sterile-active neutrino mixing parameter $\sin^{2}2\theta_{14}$ were obtained in the region of $2 \times 10^{-4}$ eV$^2 \lesssim \left|\Dm41\right| \lesssim$ 0.2 eV$^2$.

\section{Acknowledgment}
The Daya Bay experiment is supported in part by the Ministry of Science and Technology of China, the U.S. Department of Energy, the Chinese Academy of Sciences (CAS), 
the National Natural Science Foundation of China,  the New Cornerstone Science Foundation, the Guangdong provincial government, the Shenzhen municipal government, 
the China General Nuclear Power Group, the Research Grants Council of the Hong Kong Special Administrative Region of China, the Ministry of Education in Taiwan, 
the U.S. National Science Foundation, the Ministry of Education, Youth, and Sports of the Czech Republic, the Charles University Research Centre (UNCE), 
the Joint Institute of Nuclear Research in Dubna, Russia, and the National Commission of Scientific and Technological Research of Chile. 
We acknowledge Yellow River Engineering Consulting Co., Ltd., and China Railway 15th Bureau Group Co., Ltd., for building the underground laboratory. 
We are grateful for the ongoing cooperation from the China Guangdong Nuclear Power Group and China Light \& Power Company.

\bibliography{references.bib}

\end{document}